\documentclass[iop]{emulateapj}
\usepackage{soul}
\usepackage{amsmath,amsfonts,amssymb,mathrsfs,graphicx}
\usepackage[squaren]{SIunits}
\usepackage{natbib}
\usepackage{multirow}
\usepackage{hyperref}

\begin{document}
\title{Implication of the non-detection of neutrinos above 2PeV}

\author{Lee Yacobi}

\affiliation{Department of Physics, Technion, Israel}

\author{Dafne Guetta}
\affiliation{Osservatorio astronomico di Roma, v. Frascati 33, 00040 Monte Porzio Catone, Italy}
\affiliation{Department of Physics
	Optical Engineering, ORT Braude, P.O. Box 78, Carmiel, Israel}
\affiliation{Department of Physics, Technion, Israel}

\author{Ehud Behar}
\affiliation{Department of Physics, Technion, Israel}

\begin{abstract}

The IceCube telescope has detected diffuse neutrino emission, 20 events of which were reported to be above 60~TeV.
In this paper, we fit the diffuse neutrino spectrum using Poisson statistics, which are the most appropriate for the low counts per energy bin. We extend the fitted energy range and exploit the fact that no neutrinos were detected above 2~PeV, despite the high detector sensitivity around the Glashow resonance at 6.3\,PeV and beyond.
A best-fit power-law slope of $\alpha=2.9\pm 0.3$ is found with no evidence for a high-energy cutoff.
This slope is steeper than $\alpha=2.3\pm 0.3$ found by the IceCube team using a different fitting method. 
Such a steep spectrum facilitates the identification of high energy ($\gg$ PeV) neutrinos, if detected, to be due to the GZK effect of cosmic-ray protons interacting with the Extragalactic Background Light. 
We use the ratio of EeV to PeV neutrinos in GZK models to show that the currently detected PeV neutrinos could not be due to the GZK effect, because this would imply many more higher-energy neutrinos that should have been detected, but were not.
The non-detection of GZK neutrinos by IceCube despite more than essentially 1200 observing days, has already ruled out (at 95\% confidence) models that predict rates of $\sim1$ neutrino/yr or more.
We use this non-detection to quantify the confidence at which GZK models are ruled out, and compute the
additional IceCube and (in the future) ARA observing time that would rule them out with 95\% confidence if no detection is made.

\end{abstract}

\maketitle

\flushbottom


\section{Introduction}
IceCube, the km$^2$ neutrino detector in the south pole has detected 20 diffuse neutrinos above 60\,TeV with the highest energy being $\sim2\pm 0.25$\,PeV. These data were fitted by the IceCube Collaboration in the interval of 60\,TeV $<E<$ 3\,PeV to a single power-law with a normalization of $1.5\times 10^{-8}$GeVcm$^{-2}$s$^{-1}$sr$^{-1} $ at 100\,TeV, and a power law slope of $2.3\pm0.3$ \citep{IC_prl}. Steeper spectral slopes of 2.4-2.6 are obtained \citep{IC2015a,IC2015b}, when including neutrinos down to 25\,TeV in the fit, and subtracting the atmospheric-neutrino model.

The effective area of IceCube has  a sharp peak around the Glashow Resonance at 6.3\,PeV \citep{Glashow}. This makes IceCube particularly sensitive above the highest-detected neutrino energy. Moreover, the extremely-high-energy neutrino search method performed by the IceCube collaboration \citep{IC2010EHE,IC2011EHE,IC2013EHE} indicates overall increase of effective area with energy $\propto E^{0.6}$. Nevertheless this search has not yet found any neutrino above 2\,PeV.

Theoretical models predict ultra high energy neutrinos, especially due to the Greisen Ztsepin Kuzmin effect \citep[GZK,][]{GofGZK,ZKofGZK}.
The GZK effect is the photo-hadron interaction between the Ultra-High-Energy Cosmic-Ray (UHECR) protons and the Extragalactic Background Light (EBL) photons.
This interaction with the Cosmic Microwave Background (CMB, the low frequency end of the EBL) produces a cutoff in the UHECR spectrum at $\sim5\times 10^{19}$eV that was observed by \citet{HiRes} and \citet{Auger}.

According to the GZK photo-hadron decay scheme, neutrinos are produced with energy $\sim 5\%$ of the seed proton.
The resulting ultra-high-energy neutrinos escape from the interaction zone (i.e., neutrino source).
The flux of the GZK neutrinos strongly depends on the cosmic-ray composition and cosmological source evolution, which is why different models differ greatly in their neutrino flux prediction.
While the IceCube non-detection of GZK neutrinos so far has already constrained many theoretical models \citep{IC2013EHE}, future neutrino observatories, and in particular the Askaryan Radio Array (ARA) is being build with the main goal of detecting these high-energy neutrinos.
In the present work, we aim to quantify the confidence at which model families can be rejected, to anticipate the ARA neutrino detection rates, and to calculate the observing time without detection that would rule out the models with high confidence.

This paper is organized as follows: In Sec.~2 we present fits of different spectral models to the existing neutrino spectrum. Sec.~3 ascribes statistical validity to GZK models from the literature. Sec.~4 includes our discussion and conclusions.

\section{Spectral Analysis}\label{RES1}

\subsection{Method}
\label{sec:method}
We aim to fit the IceCube neutrino spectrum above 60\,TeV with several models and in particular to exploit the fact that high energy neutrinos above 2\,PeV were not detected.
All neutrinos are assumed to be the genuine diffuse astrophysical signal, after all background counts have been subtracted. The neutrino number flux density per solid angle can be defined as:

\begin{equation} \label{Phi}
\Phi=\frac{dN}{dE dt dA d\Omega}
\end{equation}

We use a forward-folding algorithm to fit two types of models: a power law 

\begin{equation} \label{PL}
E^2\Phi=\Phi_0\left(\frac{E}{100\text{TeV}}\right) ^{2-\alpha}
\end{equation}

\noindent and a power law with a cutoff at $E_c$
\begin{equation} \label{PLCO}
E^2\Phi=\Phi_0\left(\frac{E}{\rm 100TeV}\right)^{2-\alpha}\exp\left( \frac{ {\rm 100TeV}-E}{E_c} \right)
\end{equation}

\noindent We multiply each model $\Phi$ with the IceCube effective area $A^\text{eff}$, which is averaged over 4$\pi$ sr \citep{IC_prl}. We integrate over energy and multiply by the exposure time $\Delta t$ and the 4$\pi$ sr field of view to obtain the model neutrino counts $N$

\begin{equation} \label{eq_N}
N=4\pi \Delta t \int{\Phi A^\text{eff}dE}
\end{equation}

\noindent which can then be compared with the observed spectrum.

The energy range we use is from 60TeV to 10PeV, in order to include the Glashow resonance at 6.3PeV in the IceCube effective area.
We use Cash (C-)statistics \citep{Cash}, which is commonly used in astrophysics for low count rate data that is distributed according to Poisson statistics.
The fitting procedure minimizes the C-statistic computed when comparing the observed spectrum with the folded model \citep{XspecStat}.
This minimization does not take into account the data errors, which makes it suitable for low count rate (small error) data.
This method allows to readily fit unbinned data, i.e. bins with one or zero counts \citep{Cash}.
We test different binning schemes from 0.01 dex to 0.2 dex, results of which are discussed below.

\subsection{Results}

The best-fit parameters of the power law with and without a cutoff are presented in Table \ref{fit_tab}.
The Cash statistic is also listed in the last column of Table~\ref{fit_tab} to give an idea of the relative goodness of fit.

\begin{table}

	\centering
	\begin{tabular}{l c c c r}
		\hline \hline   
		Model           & \multirow{2}{*}{$\Phi_0 \left[\frac{10^{-8}\text{GeV}}{\text{cm}^2 \text{ s sr}}\right]$}  & $\alpha$  & \multirow{2}{*}{$\log\left(\frac{E_c}{\text{GeV}}\right)$}  & Cash \\ 
		&                                                                                            &           &                                                             & Rank \\
		
		\hline
		
		\citet{IC_prl}         & 1.5                 & $2.3\pm 0.3$        &                 & 57.8 \\
		Power Law (PL)         & $2.2_{-0.7}^{+0.9}$ & $2.9_{-0.3}^{+0.3}$ &                 & 50.2 \\
		PL w/cutoff            & $2.2_{-0.9}^{+1.2}$ & $2.9_{-0.4}^{+0.6}$ & $>6$            & 50.2 \\
		$\alpha$=2 PL w/cutoff & $1.9_{-0.5}^{+0.6}$ & 2                   & $5.9\pm 0.3$    & 56   \\
		
		\\
		
		$E<1$PeV dataset       & $2.3_{-0.8}^{+1}$   & $3.6_{-0.6}^{+0.7}$ &                 & 32  \\
		\hline
	\end{tabular}
	
	\caption{Fitting parameters for the three models and IceCube results. The last row is a power-law model fitted to the data, but only up to 1~PeV.}
	\label{fit_tab}
	
\end{table}

We find a slope of $\alpha = 2.9\pm0.3$, which is steeper by  2$\sigma$ than the slope of $2.3\pm 0.3$ found by \citet{IC_prl}, who use a different statistical fitting method. 
\citet{IC_prl} unfolded the observed spectrum, assuming $\alpha = 2.0$ in each bin, and then fitted the unfolded spectrum, while we directly forward-fitted the model to the count spectrum (Sec.~\ref{sec:method}).
The discrepancy may also be partially attributed to the different energy range of 60\,TeV -- 3\,PeV used by \citet{IC_prl}, which does not include the Glashow resonance.
When we fix the slope to be 2.3, and fit for the normalization only up to 3\,PeV, 
we find $\Phi _0=1.4\pm0.6$, which is indeed consistent with the \citet{IC_prl} value of 1.5 (no uncertainty quoted).

We find no evidence for a cutoff in the data. 
Fitting for a cutoff results in $E_c >$1EeV, which is far outside the fitted data range.
The statistical uncertainty on $E_c$, however, is large; the fit only restricts $E_c$ to be $>1$\,PeV.
In Table~\ref{fit_tab}, we also show results for a power law and a cutoff, with the canonical slope of $\alpha = 2$ that might be motivated by theoretical cosmic-ray acceleration processes. In this case, the best-fit cutoff energy is at $\sim$1PeV, but the fit is much worse.

It is evident from the IceCube data that no neutrinos were detected between 400\,TeV and 1\,PeV, with three neutrinos detected between 1--2 PeV.
It has been suggested that the 17 neutrinos with $E<$ 400\,TeV and the three PeV neutrinos have different origins, e.g., dark matter annihilation \citep{DMDM}.
If we ignore the three PeV neutrinos and fit a power law model.
the resulting slope of $\alpha = 3.6 \pm 0.7$ is much steeper than the slope obtained with the PeV neutrinos.
The C-statistic is obviously much better as the PeV events, which are the least consistent with a power law, were removed.

\begin{figure}[!]
	\includegraphics[width=0.49\textwidth]{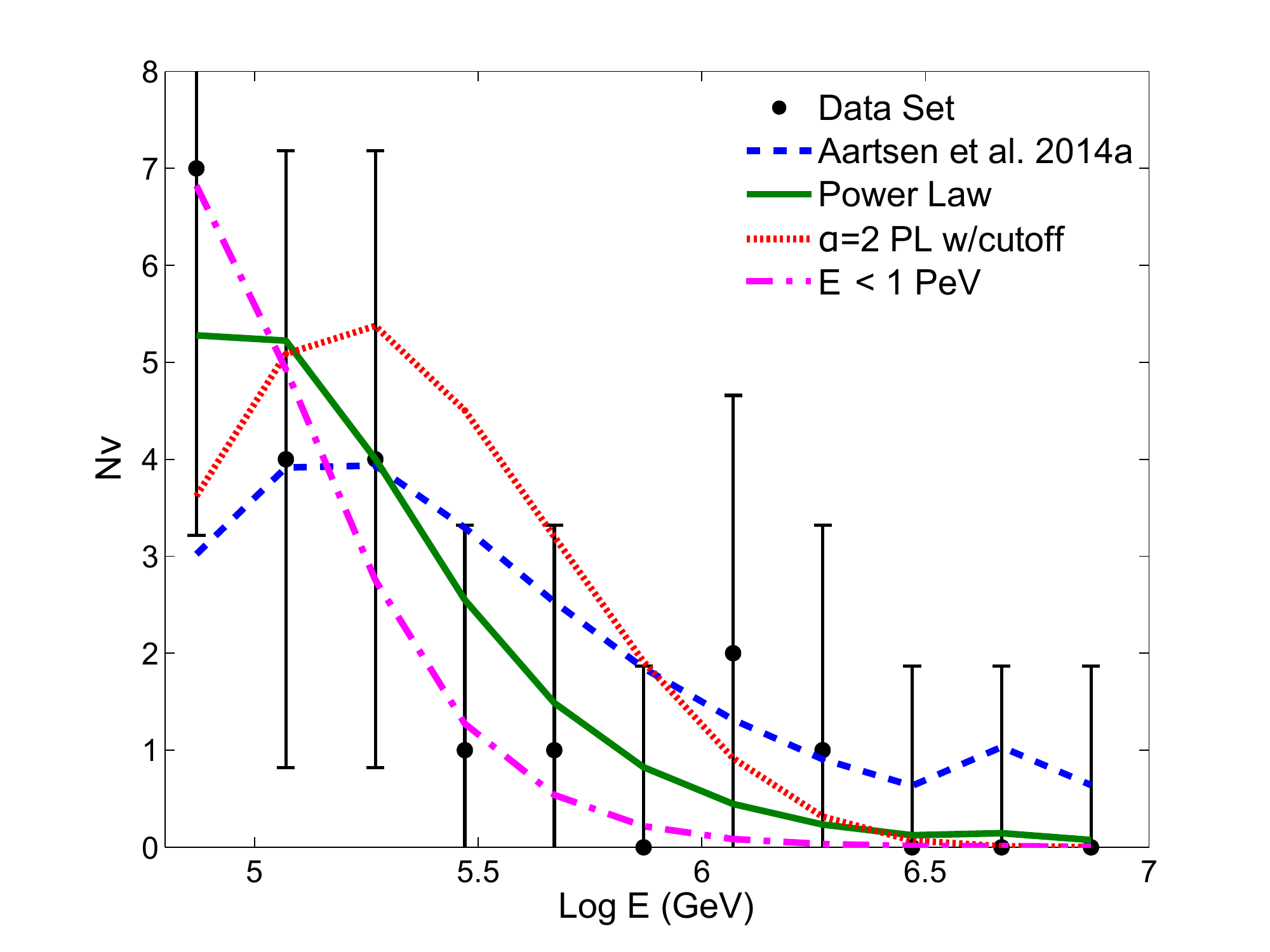}
	\caption{Observed neutrino spectrum compared with the various models (see Table \ref{fit_tab}). Data are binned to 0.2 dex. Poissonian errors of $1+\sqrt{N+0.75}$ \citep{gehrels1986} are plotted just to guide the eye.}
	\label{powerlaw}
\end{figure}

In Fig.\,\ref{powerlaw} we plot the neutrino data as well as the various models.
It can be seen that on this plot the best-fitted power law model passes closer to the data points than any other model.
This is due to the correct treatment of zero count bins.
The fixed $\alpha = 2$ power law with a cutoff seems the least appropriate, likely implying that the diffuse neutrinos are not produced at the cosmic-ray acceleration sources, which is also consistent with the lack of a clear angular coincidence with astrophysical sources \citep{aartsen2014searches}.
The differences between the other models are still very small.

Since the neutrinos do not always deposit all of their energy in the detector, some uncertainty can be associated with their energies.
The actual neutrino energies are estimated to be $\sim15\%$ greater than the deposited energy \citep{aartsen2014Energy}.
In order to test the effect of this uncertainty on the spectral parameters, we fitted the data using different bin sizes over a large range between 0.01 dex and 0.2 dex, which correspond to 2.3\% -- 45\%, respectively, of the central energy in each bin.
We find that using coarser bins makes no difference to the best-fitted spectral slope, which varies between $\alpha = 2.87-2.93$, and is totally consistent with $2.9\pm0.3$ obtained for unbinned data (Table \ref{fit_tab}).
We conclude that the small uncertainty in neutrino energy makes no difference for constraining the spectral model parameters.

\section{GZK neutrinos}\label{RES2}

As shown in the previous section, the steep power law of the diffuse neutrinos implies a minute chance for detecting neutrinos above a few PeV.
However, neutrinos at these energies are also expected to be produced by the photo-meson interaction of the EBL with high-energy cosmic rays at the $\Delta$ resonance (the GZK effect).
In this section we examine several GZK neutrino models and calculate the number of neutrinos expected to be detected by current (IceCube) and future (ARA) neutrino telescopes.
An analysis of signals from extremely-high-energy (EHE, $E > 10$~PeV) neutrinos including those produced outside of the detector, and exploiting the scaling  of the effective area with energy approximately as $E^{0.6}$ significantly increases the sensitivity of IceCube to these EHE neutrinos \citep{IC2010EHE,IC2011EHE,IC2013EHE}. 
Next, we exploit the non-detection during previous IceCube seasons prior to the complete 86 string configuration (IC86).
In Table~\ref{seasons} we show the three IceCube seasons, which add up to 1224 IC86-equivalent observing days.
Despite this effort, no neutrino was detected above 2 PeV.
 
\begin{table}
	\begin{tabular}{l c c c c}
		\hline \hline
		Season      & \# Strings & $A^{\rm{eff}}$       & Period  & IC86 equivalent \\
		May to May  &            &  ($\%$ of IC86) & (days)  & (days)           \\ \hline
		
		2010-2013 & 79-86 & 100  & 988 & 988 \\
		2008-2009 & 40 & 50  & 335 & 167 \\
		2007-2008 & 22 & 28  & 242 & 69   \\ \hline
		Total  &    &     && 1224 \\ \hline
	\end{tabular}
	
	\caption{IceCube seasons for which the EHE analysis was performed \citep{IC2010EHE,IC2011EHE,IC2013EHE}}
	\label{seasons}
	
\end{table}

In order to test different GZK models from the literature we extracted model fluxes and similarly to Sec.~\ref{sec:method} folded them through the instrument effective area to obtain neutrino counts.  
No $\tau$ neutrinos are expected from the GZK effect. However, the neutrinos oscillate between flavors to produce an equal flux density for each of the three flavors $\Phi _i^\text{obs}$: 

\begin{equation} \label{separate_fluxes}
\Phi _i^\text{obs} = \frac{1}{3}\left( \Phi _e + \Phi _\mu \right)
\end{equation}

\noindent where $\Phi _e$ and $\Phi _\mu$ are the model electron and muon neutrino flux densities, respectively.
Because of the Glashow resonance for $\bar{\nu}_e$ in IceCube, we compute the model fluxes separately for neutrinos and anti-neutrinos. 
Hence, Eq.~\ref{separate_fluxes} holds separately for anti-neutrinos as well.
Since the flux density is the same for all flavors, we can sum the effective areas of all flavors

\begin{equation} \label{separate_areas}
A^\text{eff}=\sum_{i=e,\mu,\tau}A^\text{eff}_i
\end{equation}

\noindent and the same for anti-neutrinos. 
Finally, the prediction for the total number of neutrinos ($N_{\nu}+N_{\bar{\nu}}$) is obtained by adding the results of Eq.~\ref{eq_N} for neutrinos and anti-neutrinos.
The integration over energy in Eq.~\ref{eq_N} is carried out from $E = 2$~PeV to 1~ZeV ($10^{21}$~eV).

\subsection{GZK models}
\label{sec:GZKmodels}

The theoretical scheme of the GZK effect is an interaction of a UHECR proton with the EBL. The main channel for this interaction is the $\Delta$-resonance:

\begin{equation} \label{eq_Delta}
p+\gamma \to \Delta ^{+} \to \bigg\{
\begin{gathered}
n+\pi ^{+} \\
p+\pi ^{0} \\
\end{gathered}
\end{equation}

\noindent In the case of protons in heavier nuclei, the efficiency of the interaction is reduced.
The neutrino emission is due to pion decay

\begin{equation}
\label{eq_pi_decay}
\pi ^{+}\to e^{+}+\nu_{e}+\nu _{\mu }+\bar{\nu }_{\mu }
\end{equation}

\noindent 
The resonance sets a requirement on $E_pE_\gamma$, so the proton energy required for the interaction decreases with redshift. 
The CMB photons interact with high energy protons producing a peak in the neutrino flux at $E_\nu\approx 5\%E_p\approx10^9$~GeV, while shorter wavelength photons of the EBL interact with (more abundant) lower energy protons to produce neutrinos around $E_\nu\approx10^6$~GeV.
An additional minor contribution to the $\bar{\nu}_e$ flux comes from the neutron beta decay.

\begin{figure}[!]
	\includegraphics[width=0.49\textwidth]{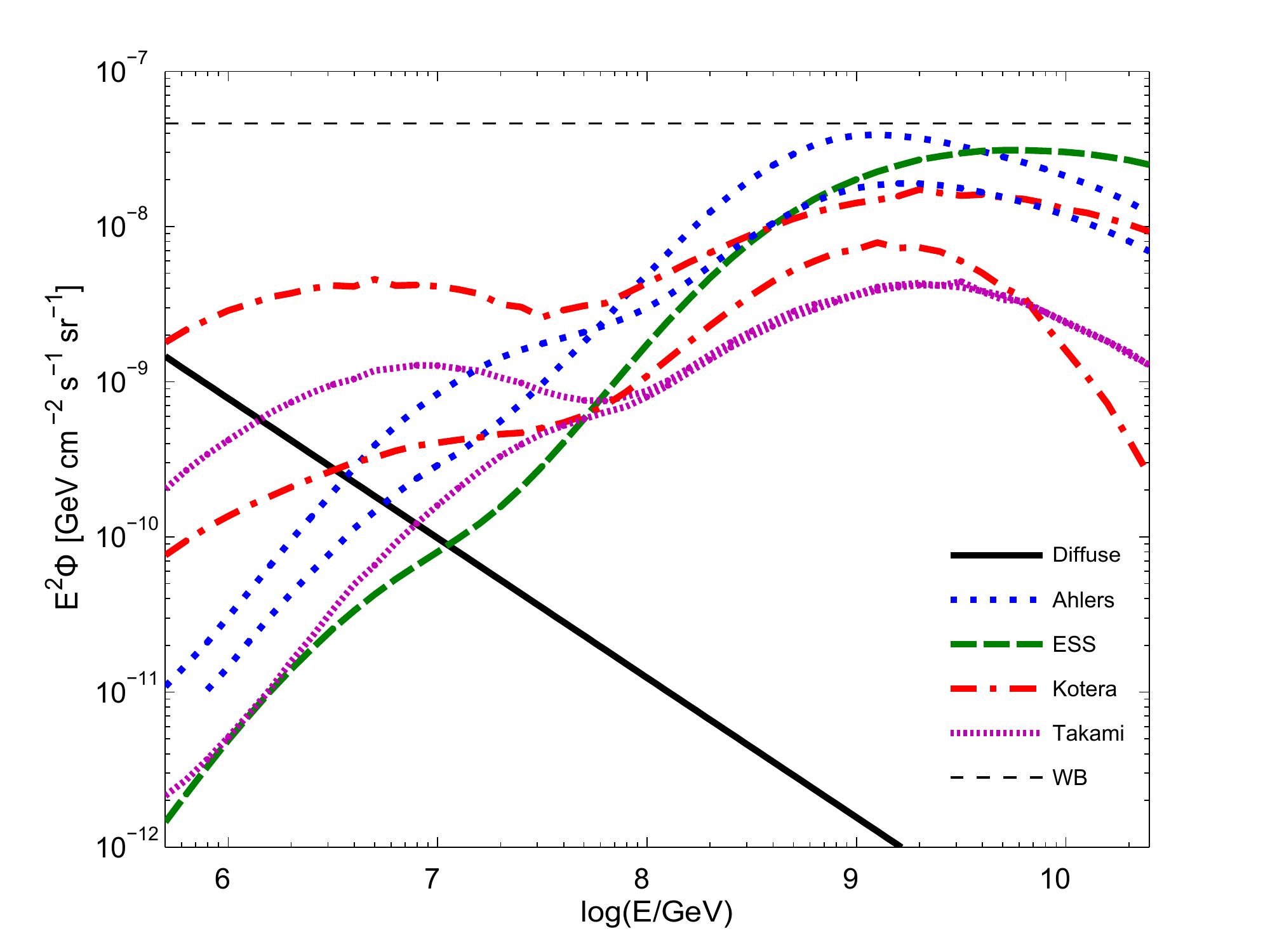}
	\caption{Various GZK neutrino models compared with the best-fitted diffuse flux model (Sec.~\ref{RES1}) and with \citet{WB99}. For each model we show the maximal and minimal predictions.}
	\label{GZK}
\end{figure}

GZK models have been constrained based on IceCube data above 100\,PeV \citep{IC2013EHE}.
We use the non-detection of neutrinos above 2\,PeV, but focus on those models that are still marginally viable.
The model families we consider here are \cite{ESS,ahlers2010gzk,kotera2010cosmogenic}; and \cite{takami2009cosmogenic}.
These models span more than three orders of magnitude in neutrino flux, which covers the predictions of many other models \citep[e.g.,][]{Protheroe1996,kalashev2002,Ave2005,Aloisio2015}.
For the most part, the models differ by their assumptions on the cosmic-ray composition, mostly the Fe content, the spectrum, and on the cosmological (redshift) evolution of the EBL and UHECR sources.
Figure \ref{GZK} shows the neutrino flux predictions of four model families from the literature. For three of the models, we show the minimal and maximal neutrino flux ($E^2\Phi$) predictions. The figure also shows for reference the best-fit diffuse spectrum of Sec.~\ref{RES1} and the upper limit of \citet{WB99}. 

The models of \citet{ESS} assume only protons and employ the SOPHIA Monte Carlo code to simulate the full particle physics interactions between UHECR and the CMB, including multi-particle products, and not only the $\Delta$ resonance. 
They also consider different cosmic ray source evolutions. In Fig.~\ref{GZK} we plot the model with the mildest evolution with redshift (their Fig.~4), which yields the least number of neutrinos. 
The fact that they include only the CMB and not shorter wavelength background results in the low prediction of neutrinos around $10^{6-7}$\,GeV.
Their other models yield many more neutrinos above 3~PeV, which have not been detected. We use their distinction between $\nu$ and $\bar{\nu}$ when considering the detection numbers for IceCube. 

The models of \citet{ahlers2010gzk} assume a pure proton cosmic-ray composition. An important parameter of this model is the energy at which the extragalactic cosmic rays dominate over the galactic component. It is the extragalactic component that produces neutrinos, and the transition energy (denoted there by $E_\text{min}$) determines the minimum proton energy for the interaction. In Fig.~\ref{GZK} we plot the predictions of \citet{ahlers2010gzk} from their Fig.~4 using the full range of $10^{17.5} - 10^{19}$~eV for this parameter.

The models of \citet{kotera2010cosmogenic} explore different cosmic-ray chemical compositions, different interacting proton energy ranges and spectra, and different cosmic-ray redshift evolution scenarios, including various transition energies between galactic and extragalactic cosmic-ray components. 
In Fig.~\ref{GZK} we plot the plausible flux range of \citet[][Fig.~9 therein]{kotera2010cosmogenic}, with parameters they consider reasonable.

\citet{takami2009cosmogenic} assume only proton cosmic rays. Here we use only their scenario in which the cosmic ray ankle at $10^{19}$\,eV is the extragalactic spectrum (dubbed there "proton dip" scenario). 
In Fig.~\ref{GZK} we plot the flux range of \citet{takami2009cosmogenic} from their Fig.~4 (left hand side) that includes minimal proton energies of $10^{7} - 10^{9}$\,GeV. 

Fig.~\ref{GZK} shows that all models predict a neutrino flux that is significantly above the diffuse spectrum at energies of $10^8 - 10^{10}$~GeV. As expected, the models of \citet{kotera2010cosmogenic} and \citet{takami2009cosmogenic} that include the IR and UV backgrounds produce a higher energy flux of neutrinos in the low-energy peak around 10$^6$~GeV.

\section{Discussion}
\subsection{Origin of PeV Neutrinos}
\begin{figure}[!]
	\includegraphics[width=0.49\textwidth]{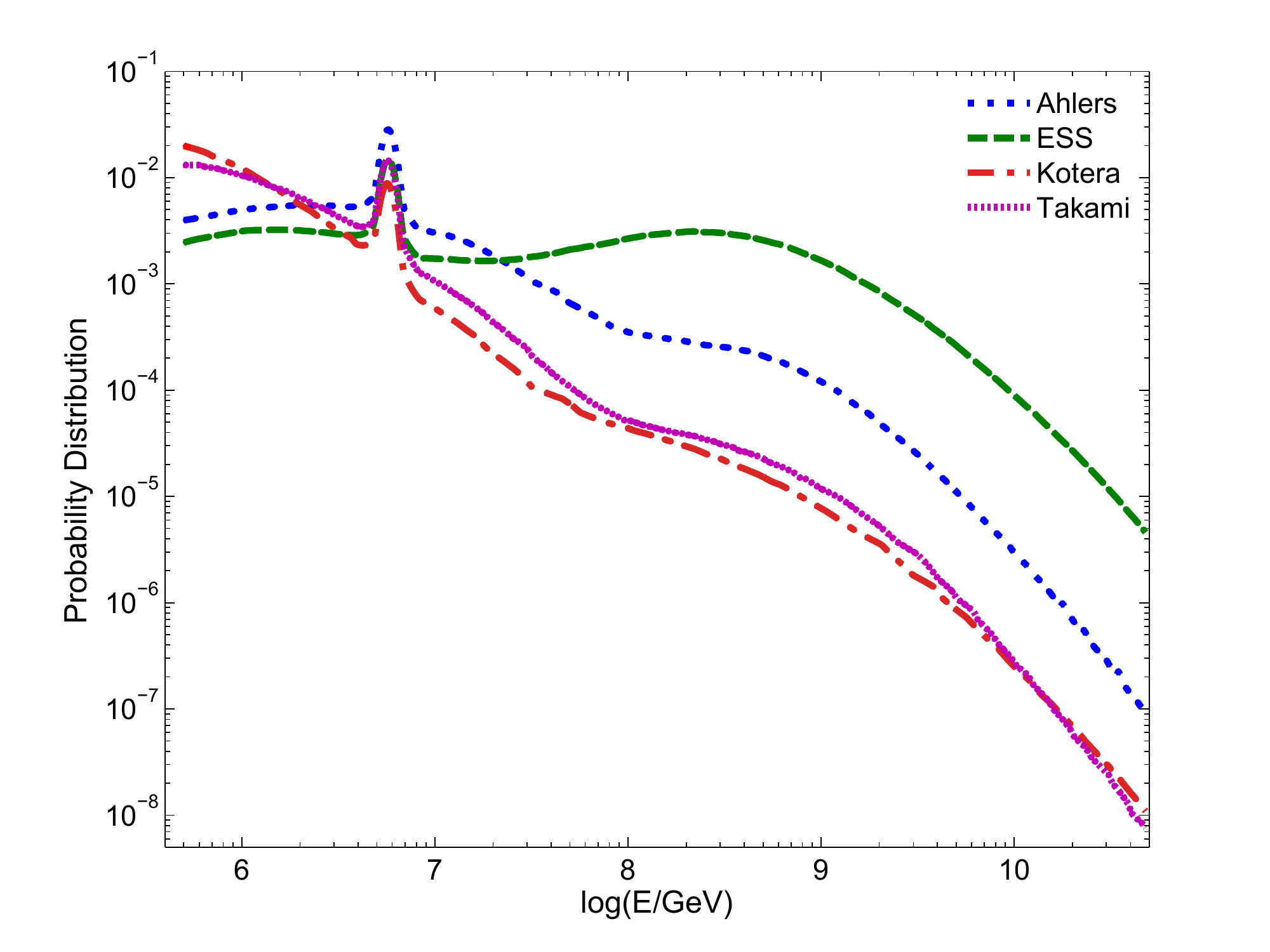}
	\caption{The neutrinos probability distribution function predicted by the various models for the IceCube effective area.}
	\label{NdE}
\end{figure}

Since the GZK models have a significant peak at a few PeV (Fig.~\ref{GZK}) that results from the EBL interaction with the UHECRs, one may wonder whether the three neutrinos detected by IceCube between 1\,PeV - 2\,PeV are actually GZK neutrinos, and not part of the diffuse power-law spectrum (Sec.~\ref{RES1}). 
In Fig.~\ref{NdE} we show the probability distribution function of neutrinos ($\propto \Phi^\text{model} A^\text{eff}$) predicted by each type of GZK model to be observed in IceCube. Evidently, in all models, the number of neutrinos predicted to be detected up to 2\,PeV is only a small fraction of the total predicted detections.
The models that predict the strongest GZK neutrino effect at a few PeV are \citet{kotera2010cosmogenic} and \citet{takami2009cosmogenic}. Even those two models predict IceCube would detect only $\approx 15\%$ of the neutrinos up to 2\,PeV, and 85\% above 2\,PeV, implying that about twenty more neutrinos should have been observed at higher energies, but were not.
This strongly suggests that the PeV neutrinos are not due to the GZK effect. 

\cite{Roulet2013} reached the same conclusion by showing that model GZK neutrino fluxes are smaller than the observed flux at PeV energies. We strengthen this argument by demonstrating that the generic spectrum of GZK neutrinos, in which the flux of CMB-produced neutrinos (EeV) is higher than that produced (at PeV energies) by longer wavelength EBL, along with the increased detection efficiency of IceCube with energy, preclude the PeV neutrinos from being due to the GZK effect, regardless of the absolute flux of any specific model.

\subsection{Constraining GZK Neutrino Models}

\begin{figure}[!]
	\includegraphics[width=0.49\textwidth]{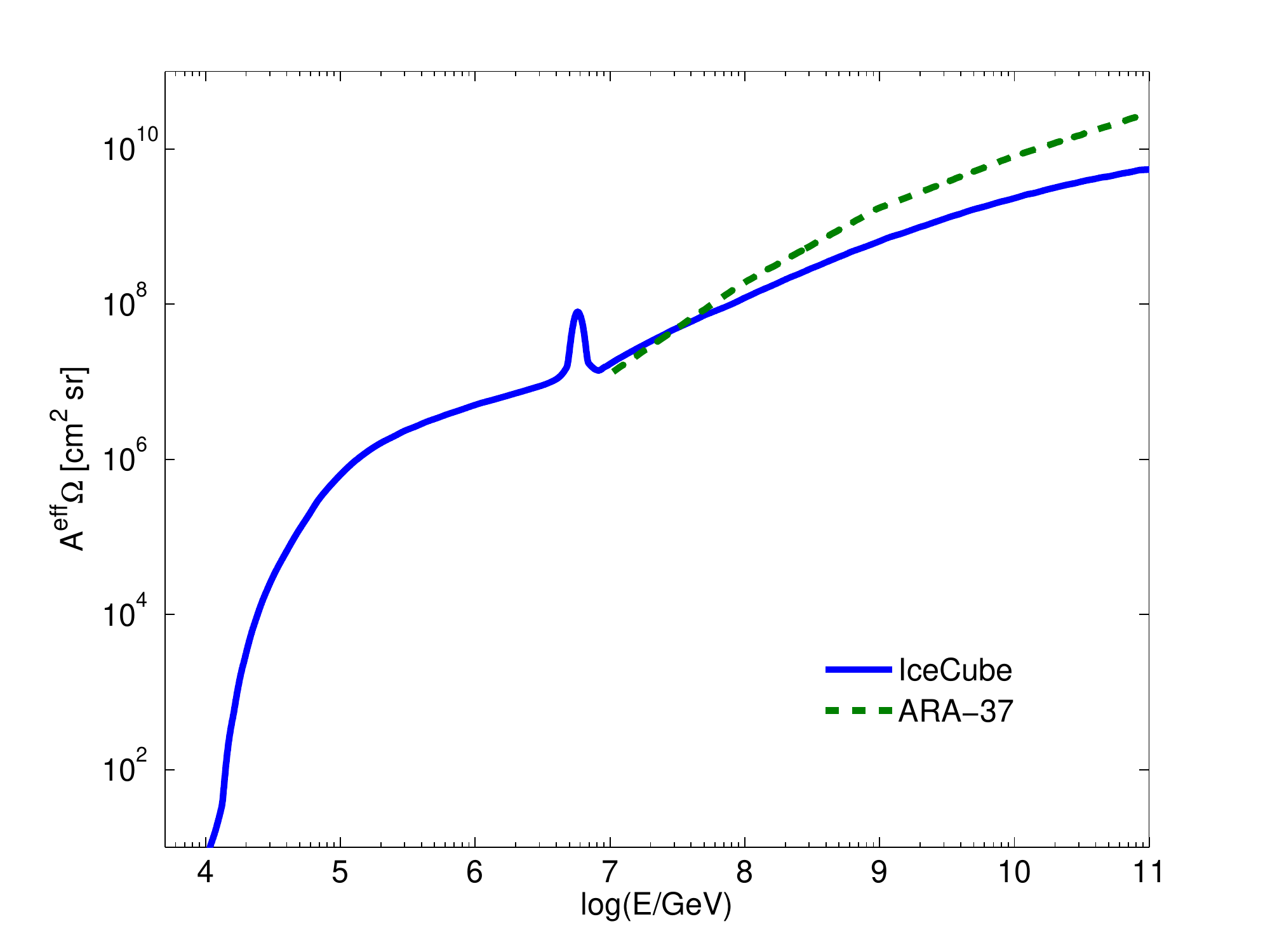}
	\caption{Effective Area times field of view of IceCube (grasp) with 86 strings and of ARA with 37 stations.
		Above the Glashow resonance, the IceCube effective area is for the EHE analysis method.}
	\label{effArea}	
\end{figure}

We test the detectability of GZK neutrinos by calculating the number of detections expected from IceCube and from ARA using Eq.~\ref{eq_N}.
In order to cover the entire energy range, we use the effective area for detecting contained neutrinos up to 10\,PeV interacting inside the detector \citep{IC_prl}, and higher energy (EHE) neutrinos, whose interaction starts outside the detector \citep{IC2013EHE}.
It was noted by \cite{karle2010} that both effective area curves match at $\sim$~30\,PeV.
The resulting effective area curve multiplied by $4\pi$ (i.e., grasp = $A^{\rm eff} \Omega$) is plotted in Fig.~\ref{effArea}.
The highest ARA curve for $A^{\rm eff} \Omega$ in \cite{ARAeff} is also plotted. 
Note that the difference in $A^{\rm eff} \Omega$ between the two telescopes is less than an order of magnitude, even at the highest energies.

The results are listed in Table~\ref{models_results}. The first column gives the number of IceCube neutrinos predicted by the various models from the actual observation time of IceCube to date. 
It can be seen that the models predict $N^\text{model} \approx 0 - 3$ detections, which allows to constrain their viability.
In the second column, we show the confidence level (CL) at which the models can be rejected given that no events were detected. The background free CL is approximately CL(\%)$ = 1 - \exp(-N^\text{model})$ \citep{astone2000upper}. 
We also compute the number of years that it would take IceCube and also ARA to reject the various models at 95\% CL. This assumes no neutrinos are detected, and 330 operational days a year.
The third column shows the additional observation time for IceCube while the fourth column lists the time for ARA.

\begin{table}
	
	\begin{tabular}{lcccc}

		\hline
		
		Model    &  IceCube          & Rejection & \multicolumn{2}{c}{Time left for 95\%}  \\
		Family   &  $N^\text{model}$ & CL (\%)   & \multicolumn{2}{c}{ CL Rejection (yrs)}    \\
		         &  (1224 days)      &           &  IceCube        & ARA  \\
		\hline
		Kotera         & 0.6-2.6    & 45-93     & 14-0.56      & 9-3      \\
		Ahlers         & 1.59-2.9   & 79-94     & 3.3-0.1      & 3.1-1.5  \\
		ESS            & 0.91       & 60        & 8.5          & 2.2      \\
		Takami         & 0.35-0.66  & 30-48     & 27-13        & 14-12.8  \\ \hline
	\end{tabular}
	\caption{GZK neutrino numbers predicted by the models, and the respective confidence levels for their rejection. Last two columns give the time, after 2013 for IceCube, and from beginning of operations for ARA, that it would take to reject the models at 95\%, given that no GZK neutrinos are detected.}
	\label{models_results}
	
\end{table}

The longer IceCube goes without detecting GZK neutrinos, the higher the statistical significance at which GZK models can be ruled out. Given that IceCube has been working for several years, all models that predicted detection rates of $\sim 1$/yr or more are already excluded with high confidence ($>95\%$).
As can be seen from Table~\ref{models_results}, the IceCube time required to seriously challenge most models is only a few years (two of which have passed).
There are still several years before neutrino detectors will be able to exclude the full range of each model family, but it is already possible to constrain the parameters inside each family. The first to be ruled out are models that include only protons and those that assume strong cosmological evolution of the UHECR sources \citep[see also][]{IC2013EHE}, which predict the highest neutrino fluxes.

\section{Conclusions}
We have studied the diffuse neutrino flux detected by IceCube. In particular, we used the non detection of neutrinos above 2 PeV to constrain diffuse neutrino spectral models and found that the best fit is obtained with a power law index of $\alpha = 2.9\pm 0.3$, which is steeper than $2.3 \pm 0.3$ found by IceCube.
More recent works extending the fitted range down to 25\,TeV already find steeper slopes \citep{IC2015a,IC2015b}.
Here we show that an extension to higher energies beyond the Glashow resonance further steepens the fitted slope.

We also use the lack of neutrino detection above 2 PeV to constrain and even reject several model families that have been suggested for neutrino fluxes expected from the GZK effect (so-called cosmogenic neutrinos). As more data are being collected, neutrino telescopes like IceCube and later ARA will allow to constrain the GZK models much better.
These observational constraints will be useful to better understand the origin of the UHECRs, the cosmological evolution of the sources, their spectrum, and chemical composition.

\acknowledgments

We are grateful for insightful discussions with M. Ahlers, H. Landsman, G. Nir, and A. Dar.
This research is supported by the I-CORE program of the Planning and Budgeting Committee and the Israel Science Foundation (grant numbers 1937/12 and 1163/10), and by a grant from Israel's Ministry of Science and Technology.
D.G. is supported by a grant from the U.S. Israel Binational Science Foundation. E.B. has received funding from the European Union’s Horizon 2020 research and innovation
programme under the Marie Sklodowska-Curie grant agreement No 655324.

\end{document}